\documentclass[hidelinks, 12pt]{article}
\usepackage[T1]{fontenc}
\usepackage{amsfonts}
\usepackage{latexsym}
\usepackage{dsfont}
\usepackage{amsmath}
\usepackage{amssymb}
\usepackage{amssymb}
\usepackage{amsthm}
\usepackage{mathrsfs}
\usepackage[nosort]{cite}
\usepackage{setspace}
\usepackage{graphicx}
\usepackage{accents}
\usepackage[font=small,labelfont=bf]{caption}
\usepackage[utf8]{inputenc}
\usepackage{geometry}
\usepackage[toc,page]{appendix}

\usepackage[hidelinks]{hyperref}
\usepackage{cleveref}
\crefformat{section}{\S#2#1#3} % see manual of cleveref, section 8.2.1
\crefformat{subsection}{\S#2#1#3}
\crefformat{subsubsection}{\S#2#1#3}

\def\appendix#1{\addtocounter{section}{1}\setcounter{equation}{0}
\renewcommand{\thesection}{\Alph{section}}
\section*{Appendix \thesection\protect\indent \parbox[t]{11.15cm}{#1}}
\addcontentsline{toc}{section}{Appendix \thesection\ \ \ #1}}

\allowdisplaybreaks
\numberwithin{equation}{section}

\def\dd{\text{d}}

\def\nt{\tilde{\nabla}}

\begin{document}

%%%%%%%%%%%%% TITLE %%%%%%%%%%%%%%%%%%%%%%%%%

\begin{titlepage}
\begin{center}
%\today
\vspace*{2.0cm}

{\LARGE  {\fontfamily{lmodern}\selectfont \bf Non-extremal near-horizon geometries}} \\[.2cm]

\vskip 2cm
{\large Andrea Fontanella} \normalsize\\
\vskip 0.5cm

\textit{Perimeter Institute for Theoretical Physics, \\
Waterloo, Ontario, N2L 2Y5, Canada} \\
\vspace{1mm}
\href{mailto:afontanella@perimeterinstitute.ca}{\texttt{afontanella@perimeterinstitute.ca}}

\end{center}

\vskip 2 cm
\begin{abstract}
\vskip1cm

\noindent When Gaussian null coordinates are adapted to a Killing horizon, the near-horizon limit is defined by a coordinate rescaling and then by taking the regulator parameter $\varepsilon$ to be small, as a way of zooming into the horizon hypersurface.   
In this coordinate setting, it is known that the metric of a non-extremal Killing horizon in the near-horizon limit is divergent, and it has been a common practice to impose extremality in order to set the divergent term to zero. Although the metric is divergent, we show for a class of Killing horizons that the vacuum Einstein's equations can be separated into a divergent and a finite part, leading to a well-defined minimal set of Einstein's equations one needs to solve. We extend the result to Einstein gravity minimally coupled to a massless scalar field. We also discuss the case of Einstein gravity coupled to a Maxwell field, in which case the separability holds if the Maxwell potential has non-vanishing components only in the directions of the horizon spatial cross section.    

\end{abstract}

\end{titlepage}

\tableofcontents
\vspace{5mm}
\hrule

%%%%%%%%%%%%%%%%%%% BODY %%%%%%%%%%%%%%%%%%

 \setcounter{section}{0}
 \setcounter{footnote}{0}

 \section*{Introduction}

Killing horizons are a particular type of null hypersurfaces which geometrically capture the event horizon of black holes \cite{Hawking:1973uf}. To approach the event horizon more closely, one can take a limit that focuses on the vicinity of the horizon itself, known as the \emph{near-horizon limit}.
The geometry obtained after taking such limit is called \emph{near-horizon geometry}. Studying near-horizon geometries find useful applications in understanding and classifying the rich structure of asymptotically flat black holes in dimensions higher than four \cite{Obers:2008pj, Emparan:2008eg}. In particular ruling out certain near-horizon geometries (e.g. AdS$_3\times S^2$) allows to exclude the existence of certain black hole solutions (e.g. a black ring \cite{Emparan:2001wn}, respectively). Near-horizon geometries of \emph{extremal} horizons, i.e. with zero temperature, have been classified in vacuum  \cite{Kunduri:2008rs} and in other theories \cite{Kunduri:2013gce}.

The problem of taking the near-horizon limit of a \emph{non-extremal} Killing horizon is that the metric becomes divergent -- an issue which does not happen if the horizon is \emph{extremal}, as the metric remains finite in the limit. This is because the metric of a non-extremal Killing horizon in Gaussian null coordinates is
\begin{equation*}
ds^2 = - r \hat{f} \dd u^2  + 2 \dd u \dd r + 2 r \hat{h}_i \dd u \dd y^i + \hat{\gamma}_{ij} \dd y^i \dd y^j
\end{equation*} 
and the near-horizon limit consists in rescaling $u \rightarrow u/\varepsilon$, $r\rightarrow \varepsilon r$, with $\varepsilon \rightarrow 0$. Assuming the metric components are analytic in $r$, extremality imposes the vanishing of the first Taylor series coefficient of $\hat{f}$, and therefore the near-horizon limit is well defined. 

In this paper we show that although in the non-extremal case the metric becomes divergent, Einstein's equations can be separated into a divergent and a finite part, and the set of equations of motion one needs to solve do not show any divergent term. This is different from a non-relativistic expansion of the metric, where one needs to consider the equations of motion at all orders in the expansion parameter \cite{VandenBleeken:2017rij}. Moreover the near-horizon geometry of a non-extremal horizon is not strictly speaking a Newton-Cartan geometry, as we shall comment in section \ref{sec:NC_geom}. 

The analysis performed here involves truncating the near-horizon metric by neglecting small correction terms of higher order in $\varepsilon$, hence our results hold for a class of near-horizon geometries where such approximation can be made. We shall further comment on this in the conclusions.   

This paper is structured as follows. In section \ref{sec:def_NHgeom} we give the definition of a Killing horizon and take the near-horizon limit in the non-extremal case. In section \ref{sec:NC_geom} we discuss whether the non-extremal Killing horizon in the near-horizon limit can be interpreted as a Newton-Cartan geometry. In section \ref{sec:Einst_vacuum} we show that the vacuum Einstein's equations for a non-extremal Killing horizon in the near-horizon limit admit a separation into divergent and finite parts, leading to a well-defined minimal set of independent equations of motion one needs to solve. In section \ref{sec:Eins_Dilaton} we extend the result of the previous section to Einstein's gravity minimally coupled to a massless scalar field. In section \ref{sec:Eins_Max} we couple Einstein's gravity to a Maxwell field and discuss under which circumstances the result found in vacuum still hold.

\section{Vacuum non-extremal horizons}\label{sec:vacuum}

The theory considered in this section is pure Einstein gravity, with action 
\begin{equation}\label{EH_action}
S_{\text{EH}} = \int \dd^D x \sqrt{-g} \, R \ , 
\end{equation}
where $g =$ det$(g_{\mu\nu})$, with $g_{\mu\nu}$ being the metric of the $D$-dimensional spacetime, and $R$ its Ricci scalar. We set the cosmological constant to zero, i.e. $\Lambda = 0$. In section \ref{sec:EMD_theory} we shall include matter fields, namely a Maxwell potential and a scalar field.

\subsection{Definition of near-horizon limit}\label{sec:def_NHgeom}

Let us assume the spacetime contains a \emph{Killing horizon}, which geometrically describes the event horizon of a black hole or a stack of branes in string theory. 
A Killing horizon $\mathcal{H}$ is a hypersurface of codimension 1 which is: (i) null, and (ii) admits a Killing vector $\xi$, which is null on $\mathcal{H}$. Due to a result of Isenberg and Moncrief \cite{Moncrief:1983xua}, see also \cite{Friedrich:1998wq, Fontanella:2018gpb}, one can adapt \emph{null Gaussian coordinates}, denoted by $(u, r, y^i)$, to a neighbourhood of the horizon, such that the Killing vector $\xi$ becomes
\begin{equation}
\xi = \frac{\partial}{\partial u}\ , 
\end{equation} 
the horizon $\mathcal{H}$ is located at $r=0$, and the spacetime metric becomes 
\begin{equation}\label{Horizon_metric}
ds^2 = - r \hat{f} \dd u^2  + 2 \dd u \dd r + 2 r \hat{h}_i \dd u \dd y^i + \hat{\gamma}_{ij} \dd y^i \dd y^j \ , 
\end{equation}
where $\hat{f}, \hat{h}_i, \hat{\gamma}_{ij}$ are a scalar, a 1-form  and a symmetric rank 2 tensor respectively, which depend only on $(r, y^i)$, due to the Killing symmetry $\xi$. The indices $i, j, ...$ run from $1, ... , D-2$ and the coordinates $y^i$ span a codimension 2 manifold, denoted by $\mathcal{S}$, fully contained inside $\mathcal{H}$. The manifold $\mathcal{S}$ is called the \emph{spatial cross section} of the horizon, and in this work we shall \emph{not} make any assumption on it, with only exception in section \ref{sec:applications}.   

To proceed further, a common assumption is that the metric components are \emph{analytic} in $r$. This allows us to Taylor expand them about the horizon at $r=0$, i.e. 
\begin{equation} \label{Taylor}
	\begin{aligned}
\hat{f}(r, y) &= f(y) + r \Delta(y) + \mathcal{O}(r^2)\\
\hat{h}_i (r,y) &= h_i(y) + \mathcal{O}(r) \\
\hat{\gamma}_{ij}(r,y) &= \gamma_{ij} (y) + \mathcal{O}(r) \ , 
	\end{aligned}
\end{equation}
A quantity associated with the metric which one may compute, which will become useful later, is the surface gravity $\kappa$, defined by the general equation $\nabla_{\mu} (\xi^ {\nu}\xi_{\nu})|_{\mathcal{H}} = - 2 \kappa\, \xi_{\mu}$. Since $\xi^{\mu}\xi_{\mu} = - r \hat{f}$, we have  
\begin{equation}
\partial_{r} (\xi^{\nu} \xi_{\nu})|_{r=0} = - f(y)  \qquad \Longrightarrow \qquad \kappa = \frac{f}{2}  \ . 
\end{equation} 
If $\kappa = 0$, i.e. $f = 0$, the horizon is extremal and it has vanishing temperature, otherwise it is non-extremal and it has non-zero temperature. 

The near-horizon limit \cite{Reall:2002bh} of the metric (\ref{Horizon_metric}) is performed by rescaling the coordinates 
\begin{equation}
u \rightarrow \frac{u}{\varepsilon} \ , \qquad\qquad
r \rightarrow \varepsilon \, r \ , 
\end{equation}
and by taking the limit $\varepsilon \rightarrow 0$. Clearly the term $r f \dd u^2$ diverges in such limit, and it is a common practice to assume the horizon to be \emph{extremal}, in order to set it identically to zero.
In this work we consider $f$ to be a generic function of $y$, and we truncate the metric expansion neglecting $\mathcal{O}(\varepsilon^n)$ terms, with $n>0$, meaning  that only the metric fields considered in (\ref{Taylor}) will appear. Our analysis holds for a class of near-horizon geometries compatible with this truncation.

\subsection{Does the geometry become Newton-Cartan?}\label{sec:NC_geom}

The metric (\ref{Horizon_metric}) in the near-horizon limit takes a form which might resemble a Newton-Cartan geometry. However here we show that strictly speaking this is not the case, due to the degeneracy of the vielbein structure and the Ricci scalar behaviour.   

The near-horizon rescaling brings the spacetime metric (\ref{Horizon_metric}) in the form
\begin{equation}\label{expanded_metric}
d s^2 = - c^2 \tau_{\mu\nu} \dd x^{\mu} \dd x^{\nu} + H_{\mu\nu} \dd x^{\mu} \dd x^{\nu} + \mathcal{O}(c^{-2}) \ , 
\end{equation}
where $c \equiv \varepsilon^{-1/2}$, and 
\begin{equation} 
	\begin{aligned}
\tau_{\mu\nu} \dd x^{\mu} \dd x^{\nu} &= r f \dd u^2 \ ,\\
H_{\mu\nu} \dd x^{\mu} \dd x^{\nu} &=- r^2 \Delta \dd u^2  + 2 \dd u \dd r + 2 r h_i \dd u \dd y^i + \gamma_{ij} \dd y^i \dd y^j 
	\end{aligned}
\end{equation}
i.e. $H_{\mu\nu}$ is what one would get in the extremal case. 
In order for the metric (\ref{expanded_metric}) to be a Newton-Cartan geometry, one needs to find a Newton-Cartan data $\{ \tau_{\mu}, m_{\mu}, e_{\mu}{}^a \}$ such that 
\begin{equation} 
	\tau_{\mu}\tau_{\nu} = \tau_{\mu\nu} \ , \qquad\qquad
	- 2 \tau_{(\mu} m_{\nu)} + e_{\mu}{}^a e_{\nu}{}^b \delta_{ab} = H_{\mu\nu} \ .
\end{equation}
and pseudo-inverses $\{ \tau^{\mu}, e^{\mu}{}_a\}$ such that
\begin{equation} 
	\begin{aligned}
\tau^{\mu} \tau_{\mu} &= 1 \ , \qquad\qquad
e^{\mu}{}_a e_{\mu}{}^b = \delta_a^b \ ,  \\
\tau^{\mu} e_{\mu}{}^a = e^{\mu}{}_a \tau_{\mu} &= 0 \ , \qquad\qquad
\tau^{\mu} \tau_{\nu} + e^{\mu}{}_a e_{\nu}{}^a = \delta_{\mu}^{\nu} \ . 
	\end{aligned}
\end{equation}
This is also equivalent of demanding that the relativistic vielbein $E_{\mu}{}^A$ for the horizon metric (\ref{Horizon_metric}) expands in large $c$ as
\begin{equation}
E_{\mu}{}^0 = c \tau_{\mu} + \frac{1}{c} m_{\mu} \ , \qquad\qquad
E_{\mu}{}^{a} = e_{\mu}{}^a \ , 
\end{equation} 
and its inverse as
\begin{equation} 
	\begin{aligned}
E^{\mu}{}_0 &= \frac{1}{c} \tau^{\mu} - \frac{1}{c^3} \tau^{\mu} m_{\nu} \tau^{\nu} + \mathcal{O}(c^{-5}) \ , \\
E^{\mu}{}_a &= e^{\mu}_a - \frac{1}{c^2} \tau^{\mu} m_{\nu} e^{\nu}{}_a + \mathcal{O}(c^{-4}) \ ,
	\end{aligned} 
\end{equation}
where $A = (0, a)$, and $a= 1, ... , D-1$. However, by direct computation, one finds that the system of equations above does not admit a solution. 

Instead, the relativistic vielbein admits the following $c$-expansion 
\begin{equation}\label{E_exp_with_n}
E_{\mu}{}^0 = c \tau_{\mu} + \frac{1}{c} m_{\mu} \ , \qquad\qquad
E_{\mu}{}^1 = \frac{1}{c} n_{\mu} \ , \qquad\qquad
E_{\mu}{}^{\sf a} = e_{\mu}{}^{\sf a} \ , 
\end{equation} 
and its inverse
\begin{equation} 
	\begin{aligned}
E^{\mu}{}_0 &=  \frac{1}{c} \tau^{\mu} - \frac{1}{c^3} \tau^{\mu} m_{\nu} \tau^{\nu} + \mathcal{O}(c^{-5}) \ , \\
E^{\mu}{}_1 &= c n^{\mu} - \frac{1}{c} \tau^{\mu} m_{\nu} n^{\nu} +  \mathcal{O}(c^{-3})\\
E^{\mu}{}_{\sf a} &= e^{\mu}{}_{\sf a}  - \frac{1}{c^2} \tau^{\mu} m_{\nu} e^{\nu}{}_{\sf a} +  \mathcal{O}(c^{-4})\ , 
			\end{aligned} 
\end{equation}
where ${\sf a} = 2, ... , D-1$. The vielbeine $\{\tau_{\mu}, m_{\mu}, n_{\mu} , e_{\mu}{}^{\sf a}\}$ and their pseudo-inverses $\{\tau^{\mu}, n^{\mu} , e^{\mu}{}_{\sf a}\}$ must satisfy the conditions
\begin{equation} 
	\begin{aligned}
\tau^{\mu} \tau_{\mu} &= n^{\mu} n_{\mu} = 1 \ , \qquad
e^{\mu}{}_a e_{\mu}{}^b = \delta_a^b \ , \qquad\qquad\quad
\tau^{\mu} n_{\mu} = n^{\mu} \tau_{\mu} = 0 \ , \\
\tau^{\mu} e_{\mu}{}^{\sf a} &=n^{\mu} e_{\mu}{}^{\sf a} = 0 \ , \qquad
e^{\mu}{}_{\sf a} \tau_{\mu} = e^{\mu}{}_{\sf a} n_{\mu} = 0 \ , \qquad 
\tau^{\mu} \tau_{\nu} + n^{\mu} n_{\nu} + e^{\mu}{}_a e_{\nu}{}^a = \delta_{\mu}^{\nu} \ . 
	\end{aligned} 
\end{equation}
A solution to the above system of equations is
\begin{equation} 
	\begin{aligned}\label{solution_with_n}
	\tau_{\mu} \dd x^{\mu} &= \sqrt{r f} \dd u  \ , \\
	m_{\mu} \dd x^{\mu} &= \frac{r^{3/2} \Delta}{2 \sqrt{f}} \dd u - \frac{\dd r}{\sqrt{r f}}  - \sqrt{\frac{r}{f}} h_i \dd y^i \ , \\
	n_{\mu} \dd x^{\mu} &= - \frac{r^{3/2} \Delta}{2 \sqrt{f}} \dd u + \frac{\dd r}{\sqrt{r f}} + \sqrt{\frac{r}{f}} h_i \dd y^i \ ,  \\
			e_{\mu}{}^{\sf a} &= \theta_{i}{}^{\sf a} \delta_{\mu}^i\ , 
		\end{aligned} 
\end{equation}
where $\theta_{\mu}{}^{\sf a}$ is the vielbein for the spatial cross section metric $\gamma_{ij}$, and their pseudo-inverses have non-vanishing components
 \begin{equation} 
	\begin{aligned}\label{inverse_solution_with_n}
\tau^{u} = \frac{1}{\sqrt{r f}} \ , \qquad
\tau^{r} = \frac{r^{3/2} \Delta}{2 \sqrt{f}} \ , \qquad
n^{r} = \sqrt{r f} \ , \qquad
e^{r}{}_{\sf a} = - r h_i \theta^i{}_{\sf a}  \ , \qquad
e^{i}{}_{\sf a} = \theta^i{}_{\sf a} \ . 
		\end{aligned} 
\end{equation}
This shows that one can engineer a $c$-expansion of the relativistic vielbein that leads to the metric (\ref{expanded_metric}), and a solution is provided in (\ref{E_exp_with_n}), (\ref{solution_with_n}), (\ref{inverse_solution_with_n}). However such solution introduces a new structure in the vielbein expansion, since the component $E_{\mu}{}^1$ is required to be subleading in large $c$. This is in  contrast with the definition of Newton-Cartan geometry, where $E_{\mu}{}^1$ is finite at leading order.

For a Newton-Cartan geometry, the Ricci scalar expands in powers of $c$ as 
\begin{equation}
R = c^2 R^{(2)} + R^{(0)} + \mathcal{O}(c^{-2}) \ , 
\end{equation}
where the divergent term is 
\begin{equation}
R^{(2)} = h^{\mu\nu} h^{\rho\sigma} \partial_{[\mu} \tau_{\rho]} \partial_{[\nu} \tau_{\sigma]} \ , \qquad\qquad
h^{\mu\nu} \equiv e^{\mu}{}_a e^{\nu}{}_b \delta^{ab} \ .
\end{equation}
This makes the Einstein-Hilbert action (\ref{EH_action}) divergent in the limit $c\rightarrow \infty$. However, one can couple to it a Maxwell field, with action
\begin{equation}
S= S_{\text{EH}} + S_{\text{Max}} \ , \qquad\quad
S_{\text{Max}} = - \frac{1}{4} \int \dd^D x \sqrt{-g} \, F_{\mu\nu} F^{\mu\nu} \ , \qquad\quad
F_{\mu\nu} \equiv 2 \partial_{[\mu} A_{\nu]} \ , 
\end{equation}
and by fine-tuning the Maxwell field to the critical value
\begin{equation}
A_{\mu}  = c \tau_{\mu} \ ,
\end{equation}
there is an exact cancellation between the divergent terms coming from $R$ and $F^2$. The determinant of the metric is also divergent, as it can be computed by using Sylvester's determinant identity,
\begin{eqnarray}
\notag
- \lim_{c\rightarrow \infty} c^{-2} g &=& - \lim_{c\rightarrow \infty} c^{-2} \, \accentset{(D)}{\text{det}}\left[ -c^2 \tau_{\mu} \tau_{\nu} + H_{\mu\nu} + \mathcal{O}(c^{-2})\right] \\
\notag 
&=& -\lim_{c\rightarrow \infty} c^{-2} \, \accentset{(D)}{\text{det}}(H_{\mu\nu}) \accentset{(1)}{\text{det}}\left[ 1 - c^2 \tau_{\mu} H^{\mu\nu} \tau_{\nu} + \mathcal{O}(c^{-2}) \right] \\
&=& \accentset{(D)}{\text{det}}(H_{\mu\nu}) \, \tau_{\mu} H^{\mu\nu} \tau_{\nu}  \ ,
\end{eqnarray}
however it contributes with an overall $c^2$ factor common to $R$ and $F^2$, and therefore irrelevant to the dynamics. 
The cancellation just described is the particle analogue of the one that occurs for a \emph{string} Newton-Cartan geometry coupled to a B-field \cite{Bergshoeff:2021bmc}.    
Now, there is no need to invoke such cancellation for the metric (\ref{expanded_metric}), since its determinant and Ricci scalar are finite,  
\begin{equation}\label{finite_g_R}
\text{det} (g_{\mu\nu})= - \text{det} (\gamma_{ij}) \ , \qquad\qquad
R = - 2 \Delta - \frac{3}{2} h^2 + 2 \nt_i h^i + \tilde{R} \ ,
\end{equation}
and therefore the Einstein-Hilbert action is finite. Naively, one may think that because the Einstein-Hilbert action does not diverge, the equations of motion should also not diverge. 
This is not exactly the case, as it is well known that evaluating an action on a field ansatz and subsequently deriving the equations of motion is \emph{not} the same as deriving the equations of motion first and evaluating them on a field ansatz\footnote{This subtlety is also discussed in the context of classical string solutions in \cite{Fontanella:2021btt}.}. It has been understood that the later procedure is the one that gives a consistent answer. 
For non-extremal horizons in vacuum, at first sight Einstein's equations are in a mixed form of divergent and finite pieces. However, as we show in the next section, it is possible to simplify them in a way that the divergent and finite terms separate. 

\subsection{Einstein's equations}\label{sec:Einst_vacuum}

Einstein's equations in vacuum are $R_{\mu\nu} = 0$, and the Ricci tensor components for a non-extremal Killing horizon are 
\begin{eqnarray}
\notag
R_{uu} &=& -\frac{r}{2\varepsilon}  ( 2 f R_{ur} + h^i \partial_i f - \nt_i \partial^i f) + r^2 \Delta^2 + \frac{3}{2} r^2 \Delta h^2  - \frac{3}{2} r^2 h^i \partial_i \Delta \\
&-& r^2 \Delta \nt_i h^i + \frac{r^2}{2} \nt_i \partial^i \Delta + \frac{r^2}{4} (\dd h)_{ij} (\dd h)^{ij} \ ,  \\
R_{ur} &=& - \Delta - \frac{1}{2} h^2 + \frac{1}{2} \nt_i h^i \ , \\
R_{ui} &=& - \frac{1}{2 \varepsilon} \partial_i f - \frac{r}{2} h^2 h_i - r (\dd h)_{ij} h^j - r \partial_i \Delta + \frac{r}{2} h_i \nt_j h^j - \frac{r}{2} \nt_j (\dd h)^j{}_i \ , \\
R_{ij} &=& \nt_{(i}h_{j)} - \frac{1}{2} h_i h_j + \tilde{R}_{ij} \ , \\
R_{rr}&=& R_{ri} = 0 \ . 
\end{eqnarray}
At first sight it seems that divergent terms appear in the equations of motion. However there is a better way of rewriting them. By using the contracted Bianchi identity evaluated along the internal coordinates, i.e. $\nabla_{\mu} R^{\mu}{}_i = \frac{1}{2} \nabla_i R$, we find that
\begin{equation}
\partial_i \Delta = \Delta h_i - 2h^j \nt_{[i}h_{j]} + \nt^j \nt_{[i} h_{j]} \ , 
\end{equation} 
which is the same expression one gets in the extremal case, i.e. it does not pick up any divergent contribution. Then we impose the equation of motion $R_{ur} = 0$, which algebraically fixes $\Delta$ in terms of the rest of the fields. By doing this, some Ricci tensor components simplify to
\begin{equation}
R_{uu} = -\frac{r}{2\varepsilon}  ( h^i \partial_i f - \nt_i \partial^i f) \ , \qquad\qquad
R_{ui} = - \frac{1}{2 \varepsilon} \partial_i f \ .
\end{equation}
Therefore Einstein's equations $R_{\mu\nu}=0$ for non-extremal Killing horizons in the near-horizon limit are equivalent to the following minimal set of conditions,
\begin{subequations}
\label{minimal_vacuum}
\begin{align}
\label{minimal_Eins_Delta}
 \Delta  &= - \frac{1}{2} h^2 + \frac{1}{2} \nt_i h^i \ , \\
 \label{minimal_Eins_Ricci}
 \tilde{R}_{ij} &= \frac{1}{2} h_i h_j -\nt_{(i}h_{j)}  \ , \\
 \label{non_extr_eom}
\partial_i f &= 0 \ . 
\end{align}
\end{subequations}
Equations (\ref{minimal_Eins_Delta}) and (\ref{minimal_Eins_Ricci}) are the minimal set of Einstein's equations one would get in the \emph{extremal} case \cite{Kunduri:2008rs, Chrusciel:2005pa}. Non-extremality requires to solve the additional equation (\ref{non_extr_eom}), which is easily solved by any constant $f$. However Einstein's equations do not fix the value of the constant, which can be arbitrary. Since $f$ is proportional to the surface gravity $\kappa$, having a constant $f$ is in agreement with the fact a stationary black hole has constant surface gravity on the event horizon \cite{Bardeen:1973gs}.

\subsubsection{Applications}
\label{sec:applications}

The result of the separation of divergent and finite parts of the Einstein's equations find an application in the context of classifying non-extremal near-horizon geometries. In the extremal case, a classification was done in \cite{Kunduri:2008rs,Kunduri:2013gce} and a similar analysis can be applied to the ``non-extremal'' equations (\ref{minimal_vacuum}).

A simple analysis can be done in $D=3$, where the spatial cross section $\mathcal{S}$ is 1-dimensional. In this case its Ricci tensor vanishes identically, and equation (\ref{minimal_Eins_Ricci}) becomes
\begin{equation}
\label{3D_minimal_Eins_Ricci}
h' = \frac{1}{2} h^2 \ . 
\end{equation}
Assuming that $\mathcal{S}$ has no boundary, which means $\mathcal{S}\cong S^1$, one can integrate (\ref{3D_minimal_Eins_Ricci}) over the spatial cross section and get $h=0$, implying also $\Delta = 0$.
In the small $\varepsilon$ (large $c$) limit, the spacetime geometry is non-Lorentzian and characterised by the $H_{\mu\nu}$ and $\tau_{\mu\nu}$ metrics in (\ref{expanded_metric}), which in this case describe $\mathbb{R}^{1,1}\times S^1$ and $\mathbb{R}$ respectively\footnote{More precisely, $\tau_{\mu\nu}$ describes $\mathbb{R}$ at each non-zero fixed value of $r$.}.

In $D>3$, as pointed out for the extremal case in \cite{Kunduri:2008rs,Kunduri:2013gce}, solving the near-horizon Einstein's equations in full generality is difficult. One approach consists in assuming extra symmetry in order to reduce the near-horizon Einstein's equations to a system of ODEs, and therefore solvable. For instance, in $D=4$ it suffices to assume the solution has an additional $U(1)$ symmetry, called axisymmetry, and in $D>4$ axisymmetry is replaced by $U(1)^{D-3}$-rotational symmetries. In $D=4$, there are uniqueness theorems for vacuum near-horizon geometries which were proven when the cosmological constant is vanishing, i.e. $\Lambda = 0$ \cite{Hajicek, Lewandowski:2002ua} and also when $\Lambda \neq 0$ \cite{Kunduri:2013gce, Kunduri:2008tk}. They state that any axisymmetric extremal near-horizon geometry, non-static and with compact spatial cross section, is isometric to the near-horizon geometry of the extremal Kerr or Kerr-(A)dS black hole. An analogue result holds also in $D=5$.  
It would be interesting to generalise these uniqueness theorems to \emph{non-extremal} horizons, which would involve solving the near-horizon Einstein's equations assuming sufficiently many rotational symmetries. This would require first to include the higher corrections in $\varepsilon$ to the near-horizon metric, as needed to describe the near-horizon geometry of the non-extremal Kerr black hole.

\section{Non-extremal horizons with matter fields}\label{sec:EMD_theory}

\subsection{Einstein-Dilaton theory}\label{sec:Eins_Dilaton}
In this section we extend the result obtained in vacuum by including matter: a minimally coupled massless scalar field. For this type of matter field, the simplification that decouples the divergent and finite pieces of the equations of motion still occurs. Let us consider the theory 
\begin{equation}\label{Ein_Dilaton_action}
S_{\text{ED}} = \int \dd^D x \sqrt{-g} \, \left( R - \frac{1}{2} \partial_{\mu} \hat{\phi} \partial^{\mu} \hat{\phi} \right) \ , 
\end{equation}
and take the spacetime metric to be (\ref{expanded_metric}). We shall assume that the Killing vector $\xi = \partial_u$ is also a symmetry for $\hat{\phi}$, 
\begin{equation}\label{u_iso_phi}
\mathcal{L}_{\xi} \hat{\phi} = 0  \qquad\Longrightarrow\qquad
\partial_{u} \hat{\phi} = 0 \ . 
\end{equation}
Assuming $\hat{\phi}$ is analytic in $r$, we Taylor expand it around $r=0$,
\begin{equation}
\hat{\phi}(r,y) = \phi(y) + \mathcal{O}(r) \ ,  
\end{equation}
where the higher order terms in $r$ drop out in the near-horizon limit.
The equations of motion are 
\begin{equation}
R_{\mu\nu} = \frac{1}{2} \partial_{\mu} \hat{\phi} \partial_{\nu} \hat{\phi}   \ , \qquad\qquad
\nabla_{\mu} \partial^{\mu} \hat{\phi} = 0 \ .
\end{equation}
By using (\ref{u_iso_phi}), and the fact $\partial_{r} \hat{\phi}=0$ in the near-horizon limit, the actual Einstein's equations one needs to solve are 
\begin{equation}
R_{uu} = 0 \ , \qquad R_{ur} = 0 \ , \qquad R_{ui} = 0 \ , \qquad
R_{ij} = \frac{1}{2} \partial_{i} \phi \partial_{j} \phi \ . 
\end{equation}
In addition, the scalar field equation reduces to 
\begin{equation}
\nabla_{\mu} \partial^{\mu} \hat{\phi} = \nt_i \partial^i \phi = 0 \ . 
\end{equation}
Given this set of equations, one can repeat the same procedure done in the vacuum case to eliminate the redundant equations of motion. 
In summary, the minimal set of equations to solve for a non-extremal Killing horizon in Einstein-Dilaton theory are 
\begin{subequations}
\begin{align}
\label{minimal_Delta_phi}
 \Delta  &= - \frac{1}{2} h^2 + \frac{1}{2} \nt_i h^i \ , \\
 \label{minimal_Ricci_phi}
 \tilde{R}_{ij} &= \frac{1}{2} h_i h_j -\nt_{(i}h_{j)} + \frac{1}{2} \partial_{i} \phi \partial_{j} \phi \ , \\
 \label{f'=0scalar}
\partial_i f &= 0 \ , \\
 \label{minimal_box_phi}
 \nt_i \partial^i \phi &= 0 \ .
\end{align}
\end{subequations}
We find the analogous result of the vacuum case. Equations (\ref{minimal_Delta_phi}),(\ref{minimal_Ricci_phi}),(\ref{minimal_box_phi}) are the minimal set of Einstein's and scalar field equations one would get for an extremal near-horizon geometry with a minimally coupled massless scalar field. Non-extremality gives again equation (\ref{f'=0scalar}), which is solved by a constant $f$, in agreement with the result of \cite{Bardeen:1973gs}.

\subsection{Einstein-Maxwell theory}\label{sec:Eins_Max}
The matter field considered in this section is a Maxwell potential. For a generic Maxwell field, the trick that separates the divergent and the finite terms in the equations of motion does not work in this case. However it is possible to impose some restrictions on the Maxwell field considered, such that the separation still occurs. 

We consider the theory
\begin{equation}\label{Ein_Maxwell_action}
S_{\text{EM}} = \int \dd^D x \sqrt{-g} \, \left( R - \frac{1}{4} F_{\mu\nu} F^{\mu\nu} \right) \ , 
\end{equation}
where $F_{\mu\nu} \equiv \partial_{\mu} \hat{A}_{\nu} - \partial_{\nu} \hat{A}_{\mu}$. We shall assume that the Killing vector $\xi = \partial_u$ is also a symmetry for $\hat{A}_{\mu}$, 
\begin{equation}\label{u_iso_A}
\mathcal{L}_{\xi} \hat{A}_{\mu} = 0  \qquad\Longrightarrow\qquad
\partial_{u} \hat{A}_{\mu} = 0 \ . 
\end{equation}
We assume the components of $\hat{A}_{\mu}$ are analytic in $r$, and therefore expand them around $r=0$ as
\begin{equation} 
	\begin{aligned}
\hat{A}_u (r,y) &= B_u (y) + r A_u (y) + \mathcal{O}(r^2) \ , \\ 
\hat{A}_r (r,y) &= A_r (y) + \mathcal{O}(r) \ , \\
\hat{A}_i (r,y) &= A_i(y) + \mathcal{O}(r) \ , \\
\end{aligned}
\end{equation}
In the near-horizon limit, the 1-form $\hat{A}$ gains a divergent term, 
\begin{equation}
\hat{A} = \left( \frac{1}{\varepsilon} B_u + r A_u \right) \dd u + A_i \dd y^i + \mathcal{O}(\varepsilon) \ , 
\end{equation} 
which makes the field strength component $F_{ui}$ divergent as well. To guarantee finiteness of the Maxwell action, we note that $F^2$ has divergent terms
\begin{equation}
F^2 =
F_{r\mu} F_{r\nu} g^{rr} g^{\mu\nu} + F_{ui} F_{\mu\nu} g^{u \mu} g^{i\nu} + \text{finite} \ , 
\end{equation}
and the necessary and sufficient condition to set the divergent terms to zero is to impose 
\begin{equation}\label{Maxwell_finite}
F_{ri} = 0 \ . 
\end{equation}
However, as we shall see, the above condition, which guarantees finiteness of the action, is not enough to separate the divergent and finite terms in the equations of motion. 

The equations of motion are 
\begin{equation}
R_{\mu\nu} = \frac{1}{2} F_{\mu\rho} F_{\nu\sigma} g^{\rho\sigma} - \frac{1}{4(D-2)} g_{\mu\nu} F^2 \ , \qquad\qquad
\nabla^{\mu} F_{\mu\nu} = 0 \ . 
\end{equation}
The $rr$-component of the Einstein's equations imposes the constraint
\begin{equation}
F_{r\mu} F_{r \nu} g^{\mu\nu} = 0 \ ,  
\end{equation}
which implies $F_{r i} F_{rj}\gamma^{ij} = 0$ and therefore $F_{ri}=0$, as required by (\ref{Maxwell_finite}). Then the $ri$-component of Einstein's equations is automatically satisfied. 

The $ur$-component of Einstein's equations fixes $\Delta$ algebraically in terms of the other fields, 
\begin{equation}
\Delta = - \frac{1}{2} h^2 + \frac{1}{2} \nt_i h^i - \frac{1}{2} F_{ur}^2 + \frac{1}{4(D-2)} F^2 \ ,  
\end{equation}
where we remind $F_{ur}$ and $F^2$ are finite. Then, by plugging the conditions found so far inside the $ui$-component of the Einstein's equations we get 
\begin{equation}\label{Max_ui_comp}
\frac{1}{2\varepsilon} \partial_i f = \frac{r}{2} h_i F_{ur}^2 - \frac{1}{2} F_{u \mu} F_{i\nu} g^{\mu\nu} \ , 
\end{equation}
where 
\begin{equation}
F_{u \mu} F_{i\nu} g^{\mu\nu} = F_{ur} F_{iu} g^{ur} + F_{ur} F_{ij} g^{rj} + F_{uk} F_{ij} g^{jk} \ , 
\end{equation}
which mixes divergent and finite terms. Therefore equation (\ref{Max_ui_comp}) couples divergent and finite terms. 
Moreover, the $uu$-component of the Einstein's equations gives
\begin{equation}
\frac{1}{2}\left(\frac{r}{\varepsilon} f + r^2 \Delta \right) F_{ur}^2 + \frac{r}{2 \varepsilon}(h^i \partial_i f - \nt_i \partial^i f) + \frac{1}{2} F_{u\mu} F_{u\nu} g^{\mu\nu} = 0 \ , 
\end{equation}
where
\begin{equation}
F_{u\mu} F_{u\nu} g^{\mu\nu} = F_{ur} F_{ur} g^{rr} + 2 F_{ur} F_{ui} g^{ri} + F_{ui} F_{uj} g^{ij} \ , 
\end{equation}
which has terms of order $\varepsilon^{-1}$ and $\varepsilon^{-2}$. Since there is no $\varepsilon^{-2}$ term from any other component of the Einstein's equations, it seems impossible to decouple the various orders of the equations of motion.

A possible solution is to couple the non-extremal horizon geometry just to a Maxwell field with non-zero components only in the spatial cross section, namely $\hat{A}_u = \hat{A}_r = 0$, and only $\hat{A}_i$ is non-zero. In this way only $F_{ij}$ survives, and the equations of motion decouple into divergent and finite parts. One can then repeat the similar procedure as done in the vacuum case and arrive to the minimal set of independent equations of motion,
\begin{subequations}
\begin{align}
\label{minimal_Delta_Maxwell}
 \Delta  &= - \frac{1}{2} h^2 + \frac{1}{2} \nt_i h^i + \frac{1}{4(D-2)} F^2\ , \\
 \label{minimal_Ricci_Maxwell}
 \tilde{R}_{ij} &= \frac{1}{2} h_i h_j -\nt_{(i}h_{j)} + \frac{1}{2} F_{ik} F_j{}^k  - \frac{1}{4(D-2)} \gamma_{ij} F^2 \ , \\
  \label{f'=0_Maxwell}
\partial_i f &= 0 \ , \\
  \label{Maxwell_eom_1}
 \nt^i F_{ij} &= 0 \ , \\
   \label{Maxwell_eom_2}
 (\dd h)_{ij} F^{ij} &=0 \ . 
\end{align}
\end{subequations}
From this we learn that when Einstein gravity is coupled to a Maxwell potential, with non-vanishing components only in the directions of the spatial cross section, we obtain the analogous result of the vacuum case. Equations (\ref{minimal_Delta_Maxwell}),(\ref{minimal_Ricci_Maxwell}),(\ref{Maxwell_eom_1}),(\ref{Maxwell_eom_2}) are the minimal set of Einstein's and Maxwell equations one would get for an extremal near-horizon geometry coupled to a Maxwell potential with non-vanishing components only in the directions of the spatial cross section. Equation (\ref{f'=0_Maxwell}), given by non-extremality, is solved by a constant $f$, in agreement with the result of \cite{Bardeen:1973gs}.

\section{Conclusions}

In this paper we explored the question whether it is consistent to consider the near-horizon limit of a non-extremal Killing horizon. We found a positive answer in vacuum and also when a certain type of matter fields are coupled to gravity.

Taking the near-horizon limit of a non-extremal Killing horizon implies that one has to deal with a divergent component of the metric $g_{uu} = -\frac{r}{\varepsilon} f +$ finite. It is a common practice in the literature to focus only on extremal horizons, such that the divergent term is set to zero. In this paper we showed that for a class of non-extremal Killing horizons in vacuum, the divergent and finite contributions to the Einstein's equations of motion decouple, and therefore it is not necessary to consider the equations of motion at all orders in the parameter $\varepsilon$, as done in general in \cite{VandenBleeken:2017rij}. One just needs to solve a minimal set of (finite) equations of motion, which consists in the equations of motion for an extremal horizon, plus the equation $\partial_i f = 0$, which is solved for $f =$ const. However Einstein's equations do not fix the value of such constant. 
This is in agreement with the fact a stationary black hole has constant surface gravity on the event horizon \cite{Bardeen:1973gs}. 

We extended this result by including matter fields. When Einstein gravity is minimally coupled to a massless scalar field, the divergent and finite parts of the equations of motion can be separated, like in the vacuum case. However when gravity is coupled to a \emph{generic} Maxwell field, we did not find a way to fully decouple the divergent from the finite part of the equations of motion, unless the Maxwell field is restricted to be non-vanishing only in the directions of the spatial cross section. 

In this work the spacetime metric expansion has been truncated at $\mathcal{O}(\varepsilon^0)$, hence the analysis presented holds for a class of near-horizon geometries compatible with such truncation. Any higher order term in the metric may give a contribution to the $\mathcal{O}(\varepsilon^{-1})$ and $\mathcal{O}(\varepsilon^0)$ parts of the Ricci tensor, as in the definition of Christoffel symbols any divergent and subleading term coming from the metric may combine together to produce finite results. By a preliminary analysis, the metric terms that contribute to the $\mathcal{O}(\varepsilon^{-1})$ and $\mathcal{O}(\varepsilon^0)$ parts of the Ricci tensor are the one including up to $\mathcal{O}(\varepsilon^2)$. 
It would be interesting in a future work to repeat the analysis presented here by including these higher order corrections, and perhaps use the result to investigate a non-extremal version of the uniqueness theorems discussed in section \ref{sec:applications}.

It would be interesting to consider \emph{supersymmetric} non-extremal Killing horizons \cite{Gran:2018ijr}, e.g. the heterotic horizons \cite{Gutowski:2009wm,Fontanella:2016aok}. For static black holes, it is known that when they are supersymmetric they also are  extremal, see e.g. \cite{Ortin:1996bz}. It might be possible that the Killing spinor equations are able to set $f=0$. 

Another interesting problem is about the bulk reconstruction of the non-extremal near-horizon geometry. It would be interesting to consider radial deformations of the non-extremal near-horizon geometry and check whether the radial moduli still satisfy elliptic PDEs as in \cite{Li:2015wsa, Fontanella:2016lzo}, implying finiteness of the moduli space.

\section*{Acknowledgments}
%%%%%%%%%%%%%%%%%%%%%%%%%%%%%%%%%%%%%%

I am grateful to T. Ort\'in for a useful conversation which inspired me to investigate this problem, and to J. Gutowski, T. Ort\'in and A. Tseytlin for useful comments on a draft of this work.  
Research at Perimeter Institute is supported in part by the Government of Canada through the Department of Innovation, Science and Economic Development and by the Province of Ontario through the Ministry of Colleges and Universities. I thank the Institute of Cosmos Sciences at the University of Barcelona (ICCUB) for their hospitality while part of this work was under completion.
I thank Lia for her permanent support. 

%%%%%%%%%%%% APPENDICIES %%%%%%%%%%%%%%%%%%%%%%

\setcounter{section}{0}
\setcounter{subsection}{0}
\setcounter{equation}{0}

\begin{appendices}

\section{Conventions}
\label{app:Conventions}

We take the metric in the ``mostly plus'' signature. 
The connection $\nabla$ is the spacetime Levi-Civita connection, while $\nt$ is the Levi-Civita connection of the horizon spatial cross section $\mathcal{S}$. Indices in $\mathcal{S}$ are contracted by using the spatial cross section metric $\gamma_{ij}$, i.e.
\begin{equation}
A^i B_i \equiv A_i B_j \gamma^{ij} \ . 
\end{equation}
We denote symmetrization of indices by round brackets $(...)$ and anti-symmetrization by square ones $[...]$, including a $1/p!$ for $p$ indices. 	

The metric for a non-extremal Killing horizon in the near-horizon limit has non-vanishing components
\begin{equation}
g_{uu} = - \frac{r}{\varepsilon} f - r^2 \Delta \ , \qquad
g_{ur} = 1 \ , \qquad
g_{ui} = r h_i \ , \qquad
g_{ij} = \gamma_{ij} \ , 
\end{equation} 
while its inverse is
\begin{equation}
g^{ur} = 1 \ , \qquad
g^{rr} = \frac{r}{\varepsilon} f +r^2 (h^2 + \Delta) \ , \qquad
g^{ri} = - r h^i \ , \qquad
g^{ij} = \gamma^{ij} \ . 
\end{equation}
The Christoffel symbols have non-vanishing components
\begin{eqnarray*}
\Gamma_{uu}^u &=& \frac{1}{2\varepsilon} f + r \Delta \ , \\ 
\Gamma_{uu}^i &=&-\frac{r}{2\varepsilon} f h^i + \frac{r}{2\varepsilon} \partial^i f - r^2 \Delta h^i  + \frac{r^2}{2} \partial^i \Delta \ ,\\
\Gamma_{uu}^r &=& \frac{r}{2\varepsilon^2} f^2 + \frac{3r^2}{2\varepsilon} \Delta f  + \frac{r^2}{2\varepsilon} f h^2 -\frac{r^2}{2\varepsilon}h^i \partial_i f + r^3 \Delta^2 + r^3 \Delta h^2 - \frac{r^3}{2} h^i \partial_i \Delta \ , \\
\Gamma_{ur}^r &=& - \frac{1}{2\varepsilon} f - r \Delta - \frac{r}{2} h^2 \ , \\
\Gamma_{ur}^i &=& \frac{1}{2} h^i \ , \\
\Gamma_{ui}^u &=& - \frac{1}{2} h_i \ , \\
\Gamma_{ri}^r &=& \frac{1}{2} h_i \ ,  \\
\Gamma_{ui}^r &=& - \frac{r}{2\varepsilon} f h_i -\frac{r}{2\varepsilon} \partial_i f - \frac{r^2}{2} h^2 h_i - \frac{r^2}{2} \Delta h_i - \frac{r^2}{2} (\dd h)_{ij} h^j - \frac{r^2}{2} \partial_i \Delta \ , \\
\Gamma_{ui}^j &=& \frac{r}{2} h_i h^j + \frac{r}{2} (\dd h)_i{}^j \ , \\
\Gamma_{ij}^r &=& r \partial_{(i}h_{j)} - r h_k \tilde{\Gamma}_i{}^k{}_j \ , \\
\Gamma_{ik}^j &=& \frac{1}{2} \gamma^{k\ell} (\partial_i \gamma_{\ell j} + \partial_j \gamma_{\ell i} - \partial_{\ell} \gamma_{ij} ) \equiv \tilde{\Gamma}_{ij}^k \ . 
\end{eqnarray*}
The Riemann tensor is defined as
\begin{equation}
R^{\rho}{}_{\sigma\mu\nu} = 2 \partial_{[\mu} \Gamma_{\nu]\sigma}^{\rho} + 2 \Gamma_{[\mu|\lambda|}^{\rho} \Gamma_{\nu]\sigma}^{\lambda} \ , 
\end{equation}
and the Ricci tensor $R_{\mu\nu} \equiv R^{\rho}{}_{\mu\rho\nu}$ has non vanishing components 
\begin{eqnarray*}
R_{uu} &=& -\frac{r}{2\varepsilon}  ( 2 f R_{ur} + h^i \partial_i f - \nt_i \partial^i f) + r^2 \Delta^2 + \frac{3}{2} r^2 \Delta h^2  - \frac{3}{2} r^2 h^i \partial_i \Delta \\
&-& r^2 \Delta \nt_i h^i + \frac{r^2}{2} \nt_i \partial^i \Delta + \frac{r^2}{4} (\dd h)_{ij} (\dd h)^{ij} \ ,  \\
R_{ur} &=& - \Delta - \frac{1}{2} h^2 + \frac{1}{2} \nt_i h^i \ , \\
R_{ui} &=& - \frac{1}{2 \varepsilon} \partial_i f - \frac{r}{2} h^2 h_i - r (\dd h)_{ij} h^j - r \partial_i \Delta + \frac{r}{2} h_i \nt_j h^j - \frac{r}{2} \nt_j (\dd h)^j{}_i \ , \\
R_{ij} &=& \nt_{(i}h_{j)} - \frac{1}{2} h_i h_j + \tilde{R}_{ij} \ . 
\end{eqnarray*}

\end{appendices}

%%%%%%%%%%%%%%%% BIBLIOGRAPHY %%%%%%%%%%%%%%%%

\bibliographystyle{nb}

\bibliography{Name}

%bibliography generated by nb.bst v1.01 (C) 2003-2010 Niklas Beisert
\begin{thebibliography}{10}
\ifx\href\asklfhas\newcommand{\href}[2]{#2}\fi
\ifx\arxivref\asklfhas\newcommand{\arxivref}[2]{\href{http://arxiv.org/abs/#1}{#2}}\fi
\ifx\doiref\asklfhas\newcommand{\doiref}[2]{\href{http://dx.doi.org/#1}{#2}}\fi
\raggedright
\small
\parskip 0pt

\bibitem{Hawking:1973uf}
S.~W.~Hawking and G.~F.~R.~Ellis,
\textit{``{The Large Scale Structure of Space-Time}''},
Cambridge University Press (2011).

\bibitem{Obers:2008pj}
N.~A.~Obers,
\textit{``{Black Holes in Higher-Dimensional Gravity}''},
\textsf{\doiref{10.1007/978-3-540-88460-6_6}{Lect.~Notes~Phys.~769,~211~(2009)}},
\texttt{\arxivref{0802.0519}{arxiv:0802.0519}}.

\bibitem{Emparan:2008eg}
R.~Emparan and H.~S.~Reall,
\textit{``{Black Holes in Higher Dimensions}''},
\textsf{\doiref{10.12942/lrr-2008-6}{Living~Rev.~Rel.~11,~6~(2008)}},
\texttt{\arxivref{0801.3471}{arxiv:0801.3471}}.

\bibitem{Emparan:2001wn}
R.~Emparan and H.~S.~Reall,
\textit{``{A Rotating black ring solution in five-dimensions}''},
\textsf{\doiref{10.1103/PhysRevLett.88.101101}{Phys.~Rev.~Lett.~88,~101101~(2002)}},
\texttt{\arxivref{hep-th/0110260}{hep-th/0110260}}.

\bibitem{Kunduri:2008rs}
H.~K.~Kunduri and J.~Lucietti,
\textit{``{A Classification of near-horizon geometries of extremal vacuum black
  holes}''},
\textsf{\doiref{10.1063/1.3190480}{J.~Math.~Phys.~50,~082502~(2009)}},
\texttt{\arxivref{0806.2051}{arxiv:0806.2051}}.

\bibitem{Kunduri:2013gce}
H.~K.~Kunduri and J.~Lucietti,
\textit{``{Classification of near-horizon geometries of extremal black
  holes}''},
\textsf{\doiref{10.12942/lrr-2013-8}{Living~Rev.~Rel.~16,~8~(2013)}},
\texttt{\arxivref{1306.2517}{arxiv:1306.2517}}.

\bibitem{VandenBleeken:2017rij}
D.~Van~den~Bleeken,
\textit{``{Torsional Newton\textendash{}Cartan gravity from the large c
  expansion of general relativity}''},
\textsf{\doiref{10.1088/1361-6382/aa83d4}{Class.~Quant.~Grav.~34,~185004~(2017)}},
\texttt{\arxivref{1703.03459}{arxiv:1703.03459}}.

\bibitem{Moncrief:1983xua}
V.~Moncrief and J.~Isenberg,
\textit{``{Symmetries of cosmological Cauchy horizons}''},
\textsf{\doiref{10.1007/BF01214662}{Commun.~Math.~Phys.~89,~387~(1983)}}.

\bibitem{Friedrich:1998wq}
H.~Friedrich, I.~Racz and R.~M.~Wald,
\textit{``{On the rigidity theorem for space-times with a stationary event
  horizon or a compact Cauchy horizon}''},
\textsf{\doiref{10.1007/s002200050662}{Commun.~Math.~Phys.~204,~691~(1999)}},
\texttt{\arxivref{gr-qc/9811021}{gr-qc/9811021}}.

\bibitem{Fontanella:2018gpb}
A.~Fontanella,
\textit{``{Black Horizons and Integrability in String Theory}''},
\texttt{\arxivref{1810.05434}{arxiv:1810.05434}}.

\bibitem{Reall:2002bh}
H.~S.~Reall,
\textit{``{Higher dimensional black holes and supersymmetry}''},
\textsf{\doiref{10.1103/PhysRevD.70.089902}{Phys.~Rev.~D~68,~024024~(2003)}},
\texttt{\arxivref{hep-th/0211290}{hep-th/0211290}},
[Erratum: Phys.Rev.D 70, 089902 (2004)].

\bibitem{Bergshoeff:2021bmc}
E.~A.~Bergshoeff, J.~Lahnsteiner, L.~Romano, J.~Rosseel and C.~\c{S}im\c{s}ek,
\textit{``{A non-relativistic limit of NS-NS gravity}''},
\textsf{\doiref{10.1007/JHEP06(2021)021}{JHEP~2106,~021~(2021)}},
\texttt{\arxivref{2102.06974}{arxiv:2102.06974}}.

\bibitem{Fontanella:2021btt}
A.~Fontanella and J.~M.~N.~Garc\'\i{}a,
\textit{``{Classical string solutions in non-relativistic AdS$_{5}$
  \texttimes{} S$^{5}$: closed and twisted sectors}''},
\textsf{\doiref{10.1088/1751-8121/ac4abd}{J.~Phys.~A~55,~085401~(2022)}},
\texttt{\arxivref{2109.13240}{arxiv:2109.13240}}.

\bibitem{Chrusciel:2005pa}
P.~T.~Chrusciel, H.~S.~Reall and P.~Tod,
\textit{``{On non-existence of static vacuum black holes with degenerate
  components of the event horizon}''},
\textsf{\doiref{10.1088/0264-9381/23/2/018}{Class.~Quant.~Grav.~23,~549~(2006)}},
\texttt{\arxivref{gr-qc/0512041}{gr-qc/0512041}}.

\bibitem{Bardeen:1973gs}
J.~M.~Bardeen, B.~Carter and S.~W.~Hawking,
\textit{``{The Four laws of black hole mechanics}''},
\textsf{\doiref{10.1007/BF01645742}{Commun.~Math.~Phys.~31,~161~(1973)}}.

\bibitem{Hajicek}
P.~Hájiček,
\textit{``{Three remarks on axisymmetric stationary horizons}''},
\textsf{\doiref{10.1007/BF01646202}{Commun.~Math.~Phys.~36,~305~(1974)}}.

\bibitem{Lewandowski:2002ua}
J.~Lewandowski and T.~Pawlowski,
\textit{``{Extremal isolated horizons: A Local uniqueness theorem}''},
\textsf{\doiref{10.1088/0264-9381/20/4/303}{Class.~Quant.~Grav.~20,~587~(2003)}},
\texttt{\arxivref{gr-qc/0208032}{gr-qc/0208032}}.

\bibitem{Kunduri:2008tk}
H.~K.~Kunduri and J.~Lucietti,
\textit{``{Uniqueness of near-horizon geometries of rotating extremal AdS(4)
  black holes}''},
\textsf{\doiref{10.1088/0264-9381/26/5/055019}{Class.~Quant.~Grav.~26,~055019~(2009)}},
\texttt{\arxivref{0812.1576}{arxiv:0812.1576}}.

\bibitem{Gran:2018ijr}
U.~Gran, J.~Gutowski and G.~Papadopoulos,
\textit{``{Classification, geometry and applications of supersymmetric
  backgrounds}''},
\textsf{\doiref{10.1016/j.physrep.2018.11.005}{Phys.~Rept.~794,~1~(2019)}},
\texttt{\arxivref{1808.07879}{arxiv:1808.07879}}.

\bibitem{Gutowski:2009wm}
J.~Gutowski and G.~Papadopoulos,
\textit{``{Heterotic Black Horizons}''},
\textsf{\doiref{10.1007/JHEP07(2010)011}{JHEP~1007,~011~(2010)}},
\texttt{\arxivref{0912.3472}{arxiv:0912.3472}}.

\bibitem{Fontanella:2016aok}
A.~Fontanella, J.~B.~Gutowski and G.~Papadopoulos,
\textit{``{Anomaly Corrected Heterotic Horizons}''},
\textsf{\doiref{10.1007/JHEP10(2016)121}{JHEP~1610,~121~(2016)}},
\texttt{\arxivref{1605.05635}{arxiv:1605.05635}}.

\bibitem{Ortin:1996bz}
T.~Ortin,
\textit{``{Extremality versus supersymmetry in stringy black holes}''},
\textsf{\doiref{10.1016/S0370-2693(98)00040-9}{Phys.~Lett.~B~422,~93~(1998)}},
\texttt{\arxivref{hep-th/9612142}{hep-th/9612142}}.

\bibitem{Li:2015wsa}
C.~Li and J.~Lucietti,
\textit{``{Transverse deformations of extreme horizons}''},
\textsf{\doiref{10.1088/0264-9381/33/7/075015}{Class.~Quant.~Grav.~33,~075015~(2016)}},
\texttt{\arxivref{1509.03469}{arxiv:1509.03469}}.

\bibitem{Fontanella:2016lzo}
A.~Fontanella and J.~B.~Gutowski,
\textit{``{Moduli Spaces of Transverse Deformations of Near-Horizon
  Geometries}''},
\textsf{\doiref{10.1088/1751-8121/aa6cbf}{J.~Phys.~A~50,~215202~(2017)}},
\texttt{\arxivref{1610.09949}{arxiv:1610.09949}}.

\end{thebibliography}

\end{document}